\documentclass[aps,prl,reprint,superscriptaddress,twocolumn,longbibliography]{revtex4-1}

\usepackage{amssymb}
\usepackage{amsmath}
\usepackage{graphicx}
\usepackage[colorlinks, linkcolor=blue, urlcolor=blue, anchorcolor=blue, citecolor=blue]{hyperref}
\usepackage{color}

\begin{document}


\title{Simulating the spin-boson model with a controllable reservoir in an ion trap}

\author{G.-X. Wang}
\thanks{These authors contribute equally to this work}%
\affiliation{Center for Quantum Information, Institute for Interdisciplinary Information Sciences, Tsinghua University, Beijing 100084, PR China}

\author{Y.-K. Wu}
\thanks{These authors contribute equally to this work}%
\affiliation{Center for Quantum Information, Institute for Interdisciplinary Information Sciences, Tsinghua University, Beijing 100084, PR China}
\affiliation{Hefei National Laboratory, Hefei 230088, PR China}

\author{R. Yao}
\affiliation{HYQ Co., Ltd., Beijing 100176, PR China}

\author{W.-Q. Lian}
\affiliation{HYQ Co., Ltd., Beijing 100176, PR China}

\author{Z.-J. Cheng}
\affiliation{Center for Quantum Information, Institute for Interdisciplinary Information Sciences, Tsinghua University, Beijing 100084, PR China}

\author{Y.-L. Xu}
\affiliation{Center for Quantum Information, Institute for Interdisciplinary Information Sciences, Tsinghua University, Beijing 100084, PR China}

\author{C. Zhang}
\affiliation{Center for Quantum Information, Institute for Interdisciplinary Information Sciences, Tsinghua University, Beijing 100084, PR China}

\author{Y. Jiang}
\affiliation{Center for Quantum Information, Institute for Interdisciplinary Information Sciences, Tsinghua University, Beijing 100084, PR China}

\author{Y.-Z. Xu}
\affiliation{Center for Quantum Information, Institute for Interdisciplinary Information Sciences, Tsinghua University, Beijing 100084, PR China}

\author{B.-X. Qi}
\affiliation{Center for Quantum Information, Institute for Interdisciplinary Information Sciences, Tsinghua University, Beijing 100084, PR China}

\author{P.-Y. Hou}
\affiliation{Center for Quantum Information, Institute for Interdisciplinary Information Sciences, Tsinghua University, Beijing 100084, PR China}
\affiliation{Hefei National Laboratory, Hefei 230088, PR China}

\author{Z.-C. Zhou}
\affiliation{Center for Quantum Information, Institute for Interdisciplinary Information Sciences, Tsinghua University, Beijing 100084, PR China}
\affiliation{Hefei National Laboratory, Hefei 230088, PR China}

\author{L. He}

\affiliation{Center for Quantum Information, Institute for Interdisciplinary Information Sciences, Tsinghua University, Beijing 100084, PR China}
\affiliation{Hefei National Laboratory, Hefei 230088, PR China}

\author{L.-M. Duan}

\email{lmduan@tsinghua.edu.cn}
\affiliation{Center for Quantum Information, Institute for Interdisciplinary Information Sciences, Tsinghua University, Beijing 100084, PR China}
\affiliation{Hefei National Laboratory, Hefei 230088, PR China}
\affiliation{New Cornerstone Science Laboratory, Beijing 100084, PR China}


\date{\today}

\begin{abstract}
The spin-boson model is a prototypical model for open quantum dynamics. Here we simulate the spin-boson model using a chain of trapped ions where a spin is coupled to a structured reservoir of bosonic modes. We engineer the spectral density of the reservoir by adjusting the ion number, the target ion location, the laser detuning to the phonon sidebands, and the number of frequency components in the laser, and we observe their effects on the collapse and revival of the initially encoded information. Our work demonstrates the ion trap as a powerful platform for simulating open quantum dynamics with complicated reservoir structures.
\end{abstract}

\maketitle

Quantum simulation is an important tool to understand quantum many-body physics \cite{feynman2018simulating,georgescu2014quantum,cirac2012goals} and is one of the most promising applications of the noisy intermediate-scale analogue or digital quantum computers \cite{RevModPhys.94.015004}. Typical digital quantum computers use two-level spins as the basic processing unit and encode other particles like fermions and bosons through e.g. the Jordan-Wignar transformation \cite{Jordan1928} or the standard binary encoding with a truncation in the particle number \cite{Sawaya2020,somma2003quantum}. In contrast, analogue quantum simulators can directly utilize the bosonic degrees of freedom and largely save the complexity of encoding, thus are preferable for the near-term study of hybrid systems like spin-boson coupled systems.

Spin-boson coupled models are fundamental physical models to describe the matter-light interaction or various couplings to bath in materials. For a single spin and a single bosonic mode, there are the well-known quantum Rabi model \cite{RevModPhys.91.025005} and Jaynes-Cummings model \cite{Jaynes1963}. With multiple spins coupled to the same bosonic mode, one has the Dicke model \cite{HEPP1973360,PhysRevA.7.831} and the Tavis-Cummings model \cite{PhysRev.170.379}, and if one further extends to multiple bosonic modes, they are generalized to the Rabi-Hubbard model \cite{PhysRevLett.109.053601,hwang2013largescale,zhu2013dispersive} and the Jaynes-Cummings-Hubbard model \cite{greentree2006quantum,angelakis2007photonblockadeinduced,hartmann2006strongly,rossini2007mottinsulating}. On the other hand, one can also consider a single spin coupled to multiple bosonic modes, which is known as the spin-boson model \cite{RevModPhys.59.1} and is a prototypical model to understand open quantum dynamics like the dissipation of an atom in an environment of electromagnetic field modes or of a solid-state qubit in a bath.

Within a continuum of the bosonic modes (a reservoir), the spin dynamics can be solved analytically \cite{RevModPhys.59.1} or numerically \cite{PhysRevB.50.15210,RevModPhys.80.395,PhysRevLett.105.050404,PhysRevLett.111.243602,PhysRevE.88.023303,PhysRevLett.119.143602}, and interesting phenomena like the non-Markovian revival of information \cite{PhysRevLett.103.210401,PhysRevLett.108.043603,PhysRevX.11.041043} has been observed. Experimentally, this model has been simulated in systems like superconducting circuits \cite{doi:10.1126/science.1181918,PhysRevX.11.041043} and neutral atoms or quantum dots in photonic crystals \cite{thompson2013coupling,Goban2014,PhysRevLett.113.093603}. As one of the leading platforms for quantum simulation, ion trap has realized many spin-boson coupled models \cite{lv2018quantum,cai2021observation,mei2021experimental,toyoda2013experimental,ohira2021blockade,debnath2018observation,PhysRevLett.129.140501}. Proposals for the spin-boson model have also been made in the ion trap system \cite{PhysRevA.78.010101,Lemmer_2018} but has not yet been realized. Here we simulate the spin-boson model in a chain of trapped ions, which naturally hosts a set of collective phonon modes with a configurable structure, and supports laser-induced coupling between the spin and the phonon modes. Specifically, we generate spin-phonon interaction for a target ion in a chain of up to 20 ions and demonstrate the collapse and revival of the initially encoded information. We engineer the reservoir structure by the ion number, the location of the target ion, the spin-phonon detuning, and by introducing multiple frequency components in the spin-phonon coupling, and we observe their effects on the simulated open quantum dynamics. Our work demonstrates the trapped ions as an ideal platform for simulating spin-boson coupled models and provides tools for the future study of complicated coupling patterns.

\begin{figure}[!tbp]
    \centering
	\includegraphics[width=8.4cm]{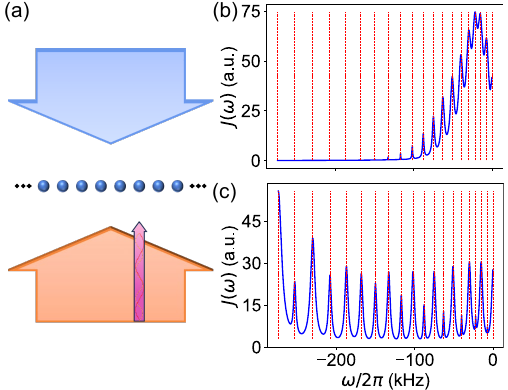}
	\caption{(a) Experimental scheme. We use counter-propagating $355\,$nm laser for Raman transitions, with a global beam for the sideband cooling of the whole ion chain and a narrow beam for the spin-phonon interaction on a selected ion. The narrow beam with a beam waist radius of $3\,\mu$m can move along the chain to address any target ion controlled by an acousto-optic deflector (AOD). Other standard lasers for the basic operations like Doppler cooling, optical pumping and qubit state detection are not shown. (b) The phonon spectrum in arbitrary unit seen by an edge ion weighted by the spin-phonon coupling strength [Eq.~(\ref{eq:density})]. Here we set a sideband Rabi rate $\eta_k \Omega=2\pi\times 6.67\,$kHz for the COM mode, which is the coupling strength we use in the following experiments. (c) A similar plot for a central ion. }\label{fig:setup}
\end{figure}
Our setup is sketched in Fig.~\ref{fig:setup}. A linear chain of $N=20$ ${}^{171}\mathrm{Yb}^+$ ions is held in a cryogenic blade trap with a transverse trap frequency $\omega_x=2\pi\times 2.397\,$MHz. We encode the spin state in the $|0\rangle\equiv|{}^2 S_{1/2}, F=0, m_F=0\rangle$ and $|1\rangle\equiv|{}^2 S_{1/2}, F=1, m_F=0\rangle$ levels of a selected ion. Due to the Coulomb interaction between the ions, their transverse oscillations couple into $N$ collective phonon modes within a band of about $\Delta\omega_x \sim e^2/(4\pi\epsilon_0 m \omega_x d^3)$ where $m$ and $e$ are the mass and the charge of the ion and $d$ is the average ion spacing.
With a pair of counter-propagating $355\,$nm global Raman laser beams, we can perform sideband cooling to all the transverse modes \cite{leibfried2003quantum}. In this experiment, due to the misalignment of the trap electrode, we have a relatively large heating rate of a few hundred quanta per second for the phonon modes close to the center-of-mass (COM) mode, thus obtaining a relatively high phonon number between 0.3 and 0.9 after sideband cooling. Nevertheless, this does not hinder our observation of the spin dynamics and may actually enhance the effective spin-phonon coupling to allow to cover more phonon modes as we will see below.

We can use a global laser beam and a counter-propagating focused laser beam to form a pair of Raman transition on the target ion near the red motional sideband and to couple its spin state with the local phonon mode, which in turn can be expanded into the $N$ collective modes. The relevant spin-boson Hamiltonian of the system thus reads
\begin{equation}
H = \frac{\Delta}{2}\sigma_{z} + \sum_{k=1}^{N}\omega_{k}a^{\dag}_{k}a_{k}+ \Omega \sum_{k=1}^{N} \eta_k b_{k} (\sigma_{+}a_{k}+\sigma_{-}a^{\dag}_{k}), \label{eq:Hamiltonian}
\end{equation}
where $\Delta$ and $\omega_k$ are the frequencies of the spin and the phonon modes in the rotating frame, and $\Omega$ is the carrier Rabi frequency of the Raman transition. $a_k$ and $a_k^\dag$ are the annihilation and creation operators of the $k$-th phonon mode with the Lamb-Dicke parameter $\eta_k$, and $b_k$ is the corresponding mode coefficient for the target ion. Note that the spin and the phonon frequencies can be shifted by the same constant without affecting any dynamics. Therefore, from now on we will use the frequency of the COM mode as the reference and set it to zero.

Regarding the phonon modes as a reservoir, the open quantum dynamics of the spin is governed by a spectral density function $J(\omega)\equiv \pi \sum_k \lambda_k^2 \delta(\omega-\omega_k)$  \cite{RevModPhys.59.1,Lemmer_2018} which takes into account the coupling strength $\lambda_k\equiv 2\eta_k b_k\Omega$ to each mode and the number of modes around each frequency $\omega$. As shown by the vertical dashed lines in Fig.~\ref{fig:setup}(b), the collective mode frequencies of a linear ion chain in a harmonic trap naturally exhibit a structured spectrum which is denser near the COM mode and is sparser in the low-frequency end. By choosing the target ion at different locations of the chain and thus different $b_k$'s, we further modulate the spectral density via the coupling strength to different modes as shown in Fig.~\ref{fig:setup}(b) and (c). Here we replace the delta function in the spectral density function by a Lorentzian lineshape to cover the power broadening \cite{leibfried2003quantum}
\begin{equation}
J(\omega) = \sum_k \frac{|2\eta_k b_k\Omega|^3/\sqrt{2}}{(\omega-\omega_k)^2+(2\eta_k b_k\Omega)^2/2}, \label{eq:density}
\end{equation}
where we set the linewidth to be $2\eta_k b_k\Omega$ which is the red-sideband coupling strength of a single mode when the phonon number is zero. In practice with a nonzero average phonon number, these peaks further broaden and we shall be able to couple to more phonon modes. Note that this spectrum is not directly used in later numerical simulation. Its purpose is just to visualize the reservoir engineering and to guide the choice of parameters in the following experiments.

\begin{figure}[!tbp]
	\centering
	\includegraphics[width=8.4cm]{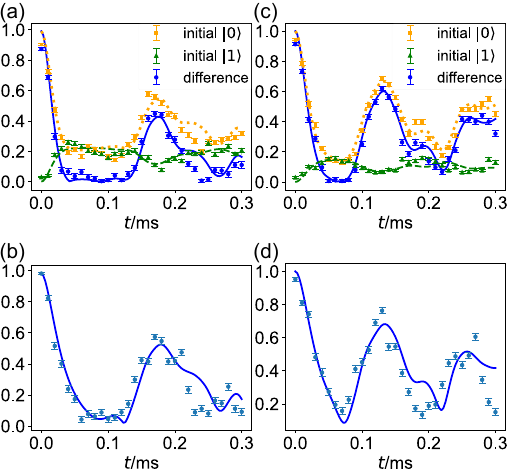}
	\caption{Non-Markovian spin dynamics. (a) Spin dynamics for an edge ion in a chain of $N=20$ ions from the initial state of $|0\rangle$ (yellow square) and $|1\rangle$ (green triangle) and their difference (blue circle). Here we tune the spin frequency to the dense part $\Delta=-2\pi\times 20\,$kHz of the phonon spectrum in Fig.~\ref{fig:setup}(b). The two states quickly become indistinguishable as the initially encoded information leaks to the phonon modes, but then the information flows backwards due to the discrete nature of the reservoir. Theoretical curves are computed by truncating to the dominant phonon modes as described in Supplemental Material. (b) A similar plot for the edge ion from the initial state of $|+\rangle$ and $|-\rangle$. Here we perform quantum state tomography for the final spin states and compute their trace distance. Similar collapse and revival are observed for the information encoded in the superposition basis. (c, d) Similar plots for an edge ion in a chain of $N=10$ ions with roughly the same ion spacing. The revival becomes more significant due to the sparser reservoir spectrum. Error bars represent one standard deviation from 300 trials.}\label{fig:nonmarkovian}
\end{figure}

First we demonstrate the non-Markovian spin dynamics in the phonon reservoir under the spin-phonon interaction. As shown in Fig.~\ref{fig:nonmarkovian}, we initialize an edge ion in orthogonal states $|0\rangle$ and $|1\rangle$ by the $355\,$nm Raman laser, and we turn on the spin-phonon coupling with a sideband Rabi rate $\eta_k\Omega=2\pi\times 6.67\,$kHz on the COM mode and a spin frequency $\Delta=-2\pi\times 20\,$kHz at the dense region of the phonon spectrum in Fig.~\ref{fig:setup}(b). We evolve the system for time $t$ and measure the final population of the spin in $|0\rangle$ as the yellow squares and the green triangles, respectively. Initially these two orthogonal states are well distinguishable up to the $2\%$ detection error and the $3\%$ state preparation error due to the axial motion of the ion and the pointing of the narrow laser. As the system evolves, the encoded information gradually leaks into the phonon reservoir and the two states can no longer be distinguished. This is reflected on Fig.~\ref{fig:nonmarkovian}(a) as the flat region near zero of the blue data points, which is the absolute difference between the populations from the two initial states. However, because of the finite and discrete phonon modes, as well as the fact that the dynamics is much faster than the motional decoherence on the time scale of milliseconds, the information in the reservoir can still flow backwards into the spin before it is eventually lost into the larger environment. This can be seen as the revival peak in the blue data points around $t_r=0.18\,$ms. We can also verify this experimental result by a numerical simulation based on the calibrated experimental parameters. Due to the exponential increase in the Hilbert space dimension under finite (thermal) phonon numbers, a direct calculation with even a moderate truncation to the phonon numbers will be challenging, so we further truncate the number of involved phonon modes based on their detuning and the relative coupling strength (see Supplemental Material for more details). As shown by the curves in Fig.~\ref{fig:nonmarkovian}, we obtain good agreement between the experimental and the theoretical results without fitting parameters, thus confirming the successful quantum simulation of the spin-boson model.

The collapse and revival of the information occur not only in the $|0\rangle$/$|1\rangle$ (particle number) basis but also in the $|+\rangle$/$|-\rangle$ (superposition) basis. In Fig.~\ref{fig:nonmarkovian}(b) we initialize the spin state in $|+\rangle$ and $|-\rangle$ by a $\pi/2$ pulse of the Raman laser. Then we repeat the above process to evolve the system under the spin-boson model Hamiltonian, and we perform quantum state tomography for the final spin states $\rho_\pm$ from the two initial states by measuring the expectation values of $\sigma_x$, $\sigma_y$ and $\sigma_z$. By definition, the maximal retrievable information can be given by their trace distance $D(\rho_+,\rho_-)$ \cite{nielsen2000quantum}. As we can see, in this case the initial decay of the information is slower, since ideally the dephasing time $T_2$ for the spontaneous emission will be twice the relaxation time $T_1$ in the particle number basis. Nevertheless, we can observe a similar flat region around zero and a revival peak at roughly the same time of $t_r=0.18\,$ms, which confirms the phase coherence of the revival signal.

Next we demonstrate our capability to manipulate the spectral density of the reservoir. One straightforward way is to adjust the ion number. Ideally, if we increase the ion number while keeping the average ion spacing unchanged, we can expect there to be more and more phonon modes in the fixed band of $\Delta\omega_x$. In Fig.~\ref{fig:nonmarkovian}(c) and (d), we compare the above results with $N=10$ ions at roughly the same ion spacing by raising the axial trap frequency. Specifically, we place the spin frequency at the same location $\Delta=-2\pi\times 20\,$kHz of the phonon spectrum. Here the initial information leakage rate does not change significantly as the decreased mode number is compensated by the increased mode coefficients $b_k$. However, due to the sparser phonon modes, we can clearly see that the revival of the information gets faster and stronger.

\begin{figure}[!tbp]
	\centering
	\includegraphics[width=8.4cm]{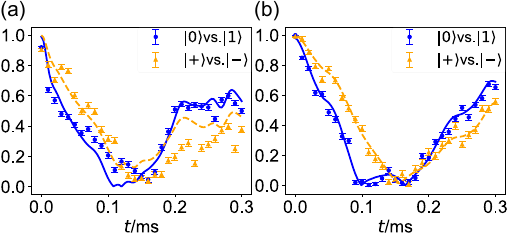}
	\caption{Reservoir engineering by a single-frequency laser. (a) By tuning the frequency of the Raman laser, we place the edge ion into a sparser region of the phonon reservoir $\Delta=-2\pi\times 50\,$kHz with weaker coupling strength. (b) By focusing the addressing laser on a central ion, we change the spin-phonon coupling pattern to Fig.~\ref{fig:setup}(c) such that the coupling gets weaker at the dense region $\Delta=-2\pi\times 15\,$kHz. In both cases, the leakage of the information into the environment slows down and the backflow of information gets much higher.  Error bars represent one standard deviation from 300 trials. }\label{fig:engineer}
\end{figure}

In practice, it is often difficult to keep increasing the ion number in a linear chain while maintaining a constant ion spacing: Due to the experimental noise, the ion number in a 1D chain is usually below 100 for a room-temperature trap \cite{zhang2017Observation,joshi2023simulation,PRXQuantum.4.010302} and below 200 for a cryogenic trap \cite{Pagano_2018,PhysRevA.106.062617}. Although it is also possible to utilize the transverse modes of a 2D ion crystal \cite{Szymanski2012crystal,PhysRevA.105.023101,PRXQuantum.4.020317,qiao2022observing,guo2023siteresolved}, it will be desirable to have more flexible ways to manipulate the structure of the phonon reservoir. In Fig.~\ref{fig:engineer}(a) we engineer the spectral density $J(\omega)$ by tuning the frequency of the spin $\Delta\to\Delta^\prime$. As we can see from Eq.~(\ref{eq:Hamiltonian}), this is equivalent to a shift in the frequency of all the phonon modes and thus a corresponding shift in the spectral density in Eq.~(\ref{eq:density}) as $J(\omega)\to J(\omega-\Delta+\Delta^\prime)$. Specifically, we set $\Delta^\prime = -2\pi\times 50\,$kHz at a sparser and weaker region of the phonon spectrum for $N=20$ ions in Fig.~\ref{fig:setup}(b). In this case, the leakage of the information into the reservoir becomes much slower. Again, the decay for the $|0\rangle$/$|1\rangle$ basis is faster than that for the $|+\rangle$/$|-\rangle$ basis, and again we observe the revival signal after a flat basin in the information.

According to Eq.~(\ref{eq:density}), another way of reservoir engineering is to place the target spin at different locations of the chain, hence modifying the mode coefficients $b_k$. In Fig.~\ref{fig:engineer}(b), we move the narrow Raman laser to a central ion of the $N=20$ chain and repeat the above experimental sequence. From the comparison between Fig.~\ref{fig:setup}(b) and (c), we can see that for the edge ion, the coupling mainly concentrates at the high-frequency end near the COM mode and the spectrum is nearly continuous in the densest region, while for the central ion the coupling is more uniform and shows a discrete feature under the same spin-phonon interaction strength. Also note that near the COM mode in Fig.~\ref{fig:setup}(c), the coupling to the odd modes are suppressed owing to the reflection symmetry of the ion chain, which effectively further increases the separation between the relevant phonon modes. Therefore, we can expect a slower leakage of the encoded information in the spin and a stronger revival, as is what we observe in the experiment.

\begin{figure}[!tbp]
	\centering
	\includegraphics[width=8.4cm]{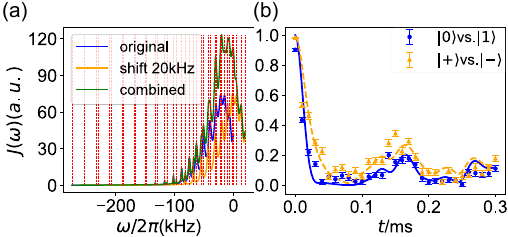}
	\caption{Reservoir engineering by a bichromatic coupling laser. (a) The phonon spectrum seen by an edge ion [blue, same as Fig.~\ref{fig:setup}(b)] and that for a laser shifted by $\delta=2\pi\times 20\,$kHz (orange). The vertical dashed lines indicate the location of the two sets of phonon modes. Since the frequency separation $\delta$ is much larger than the coupling strength $\eta_k b_k\Omega$, the two sets of spectra can add up incoherently to give the green curve. (b) At the dense region of the combined spectrum $\Delta=-2\pi\times 20\,$kHz, the decay becomes even faster than Fig.~\ref{fig:nonmarkovian} and the revival becomes less significant. Error bars represent one standard deviation from 300 trials.}\label{fig:two_freq}
\end{figure}

From the above examples, we see that the coupling of the spin to the phonon reservoir is often inefficient in the sense that only a small fraction of the phonon modes can be coupled. This is because within our achievable parameter regime, often the coupling strength $\eta_k\Omega$ below $2\pi\times 10\,$kHz is much smaller than the width of the phonon band above $2\pi\times 100\,$kHz. (To narrow down the phonon band requires larger ion distances and weaker axial trapping, making the system more sensitive to the experimental noise as we mention above.) Also we can see this from our numerical simulation method where truncation to only a few nearby phonon modes is sufficient for convergence. On the one hand, this suggests that we may need a large number of ions to approximate a continuous phonon spectrum. On the other hand, this fact also provides us with further adjustability of the spectrum by adding more frequency components in the driving laser to couple to the other unused phonon modes.

To demonstrate this idea, we choose two frequency components in the narrow Raman laser beam separated by $\delta=2\pi\times 20\,$kHz. Each frequency component generates a spin-boson Hamiltonian in the form of Eq.~(\ref{eq:Hamiltonian}), and their joint effect is just to combine the two spectra together which are separated by $\delta$, as shown in Fig.~\ref{fig:two_freq}(a). Note that strictly speaking this picture is incorrect because the two frequency components couple to the same set of phonon modes and should be treated coherently. However, so long as the coupling strength to individual phonon mode $\eta_k b_k \Omega$ is much weaker than the frequency separation $\delta$, the two frequency components will never excite the same phonon mode significantly at the same time. Then our approximate model will be valid. Here to keep consistency with the previous results, we choose the same coupling strength $\eta_k \Omega=2\pi\times 6.67\,$kHz to the COM mode for both frequency components. From the combined spectral density (green curve) in Fig.~\ref{fig:two_freq}(a), we expect denser phonon modes with stronger coupling strength. This is consistent with the experimental and numerical results in Fig.~\ref{fig:two_freq}(b) where we get faster leakage of the information and a weaker revival signal.

To sum up, we simulate a spin-boson model in a linear chain of trapped ions and study the open quantum dynamics of a spin in a structured reservoir. We demonstrate the strong controllability of the spectral density of the reservoir by various degrees of freedom including the ion number, the target ion location, the laser detuning to the phonon sidebands, and the number of frequency components in the laser. We observe the change in the collapse and revival of the encoded information under the reservoir engineering, and we confirm the experimental results with numerical simulation. Our work provides convenient tools to engineer complicated bosonic reservoir for spins, and our method can be readily generalized to a 2D ion crystal to further increase the ion number and to produce more complex coupling patterns. It is also convenient to introduce more spins into the model by adding more focused laser beams on different ions, which will allow the study of multiple emitters in a shared bosonic environment.

\begin{acknowledgments}
This work was supported by Innovation Program for Quantum Science and Technology (2021ZD0301601), Tsinghua University Initiative Scientific Research Program, and the Ministry of Education of China. L.M.D. acknowledges in addition support from the New Cornerstone Science Foundation through the New Cornerstone Investigator Program. Y.K.W. acknowledges in addition support from Tsinghua University Dushi program and the start-up fund. C.Z. acknowledges in addition support from the Tsinghua University Shuimu Scholar postdoctoral fellowship.
\end{acknowledgments}

%

\end{document}



\title{Supplemental Material for ``Simulating the spin-boson model with a controllable reservoir in an ion trap''}

\author{G.-X. Wang}
\thanks{These authors contribute equally to this work}%
\affiliation{Center for Quantum Information, Institute for Interdisciplinary Information Sciences, Tsinghua University, Beijing 100084, PR China}

\author{Y.-K. Wu}
\thanks{These authors contribute equally to this work}%
\affiliation{Center for Quantum Information, Institute for Interdisciplinary Information Sciences, Tsinghua University, Beijing 100084, PR China}
\affiliation{Hefei National Laboratory, Hefei 230088, PR China}

\author{R. Yao}
\affiliation{HYQ Co., Ltd., Beijing 100176, PR China}

\author{W.-Q. Lian}
\affiliation{HYQ Co., Ltd., Beijing 100176, PR China}

\author{Z.-J. Cheng}
\affiliation{Center for Quantum Information, Institute for Interdisciplinary Information Sciences, Tsinghua University, Beijing 100084, PR China}

\author{Y.-L. Xu}
\affiliation{Center for Quantum Information, Institute for Interdisciplinary Information Sciences, Tsinghua University, Beijing 100084, PR China}

\author{C. Zhang}
\affiliation{Center for Quantum Information, Institute for Interdisciplinary Information Sciences, Tsinghua University, Beijing 100084, PR China}

\author{Y. Jiang}
\affiliation{Center for Quantum Information, Institute for Interdisciplinary Information Sciences, Tsinghua University, Beijing 100084, PR China}

\author{Y.-Z. Xu}
\affiliation{Center for Quantum Information, Institute for Interdisciplinary Information Sciences, Tsinghua University, Beijing 100084, PR China}

\author{B.-X. Qi}
\affiliation{Center for Quantum Information, Institute for Interdisciplinary Information Sciences, Tsinghua University, Beijing 100084, PR China}

\author{P.-Y. Hou}
\affiliation{Center for Quantum Information, Institute for Interdisciplinary Information Sciences, Tsinghua University, Beijing 100084, PR China}
\affiliation{Hefei National Laboratory, Hefei 230088, PR China}

\author{Z.-C. Zhou}
\affiliation{Center for Quantum Information, Institute for Interdisciplinary Information Sciences, Tsinghua University, Beijing 100084, PR China}
\affiliation{Hefei National Laboratory, Hefei 230088, PR China}

\author{L. He}

\affiliation{Center for Quantum Information, Institute for Interdisciplinary Information Sciences, Tsinghua University, Beijing 100084, PR China}
\affiliation{Hefei National Laboratory, Hefei 230088, PR China}

\author{L.-M. Duan}

\email{lmduan@tsinghua.edu.cn}
\affiliation{Center for Quantum Information, Institute for Interdisciplinary Information Sciences, Tsinghua University, Beijing 100084, PR China}
\affiliation{Hefei National Laboratory, Hefei 230088, PR China}
\affiliation{New Cornerstone Science Laboratory, Beijing 100084, PR China}

\maketitle

\section{Experimental setup}
We trap a chain of ${}^{171}\mathrm{Yb}^{+}$ ions in a Paul trap with segmented blade electrodes and an RF frequency $\omega_{\mathrm{rf}}=2\pi\times 35.733\,$MHz. A magnetic field of $B=5.5\,$G perpendicular to the axial direction of the trap is used to define the quantization axis. We follow the standard procedure to use $369.5\,$nm laser for Doppler cooling, optical pumping and qubit state detection \cite{debnath2016programmable}, and we use counter-propagating $355\,$nm Raman laser (Paladin Compact $355$-$4000$ with a repetition rate of $118.7\,$MHz) to couple the spin and the phonon modes. As shown in Fig.~1(a) of the main text, the frequency and phase of each $355\,$nm beam can further be adjusted by an acousto-optic modulator (AOM). The $355\,$nm laser beams are oriented at $45^\circ$ to the transverse $x$ and $y$ directions. We use elliptic global Raman beams for sideband cooling of the ion chain, and an additional narrow beam with a Gaussian waist radius (where the intensity drops to $1/e^2$) of $3\mu$m$\times 5\mu$m for generating the spin-boson model and for preparing the initial state of the target spin. The orientation of the narrow beam can be controlled by an acousto-optic deflector (AOD) to select different target ions.

For a chain of $N=20$ ions, we adjust the axial confinement to achieve an average ion spacing of about $4.6\,\mu$m. We can read the ions' locations on the CCD camera with an accuracy of about $0.6\,\mu$m. To further improve the accuracy, we can use the frequencies of the transverse phonon modes to fit the ion spacing \cite{PhysRevLett.129.140501}. Then we can compute the phonon mode vectors and thus obtain $b_k$'s for the target ion.

\section{Sideband cooling}
\begin{figure}[htp]
	\centering
	\includegraphics[width=12.6cm]{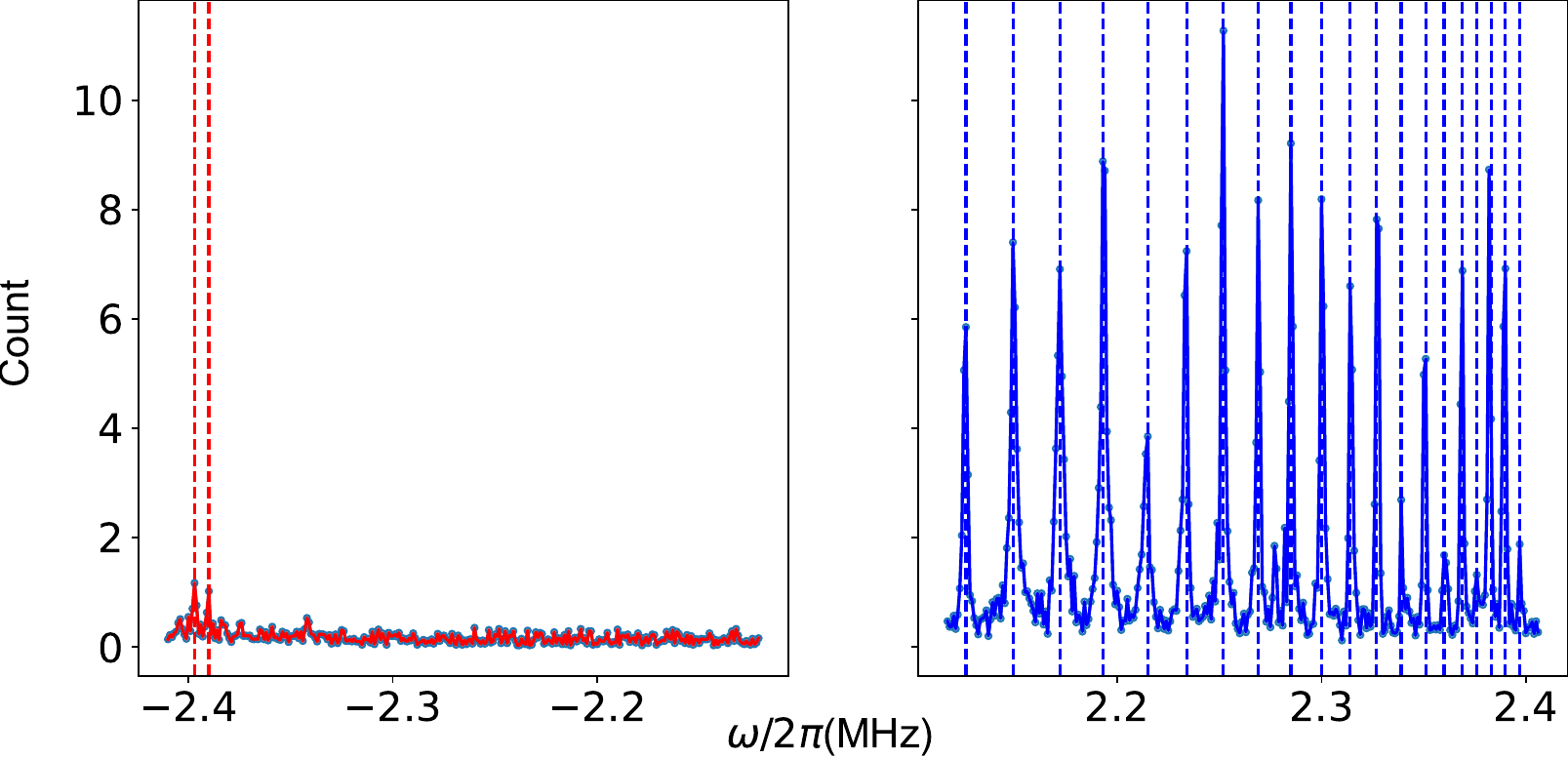}
	\caption{Spectra of the red and blue sidebands for $N=20$ ions under a weak global Raman laser. Vertical dashed lines represent the position of the phonon modes. For the red sideband, only the COM mode and the tilt mode have visible population.} \label{fig:spectrum}
\end{figure}

Before each experimental trial, we initialize the motional state by $3\,$ms Doppler cooling. Then we perform $9.75\,$ms sideband cooling for the $x$ and $y$ modes by sequentially cooling different frequencies. The spectra of the red and blue motional sidebands under the same weak driving \cite{PhysRevLett.129.140501} is shown in Fig.~\ref{fig:spectrum} for the $N=20$ ions. As we can see, apart from the center-of-mass (COM) mode and the adjacent tilt mode, all the other modes have been cooled to a low phonon number below 0.3.

\begin{figure}[htp]
	\centering
	\includegraphics[width=12.6cm]{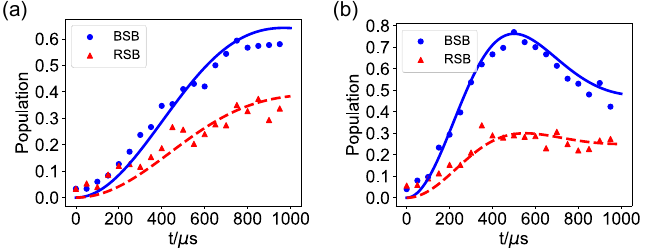}
	\caption{Blue and red sideband dynamics for the COM mode and the tilt mode with a narrow Raman laser on the edge ion.} \label{fig:phonon}
\end{figure}

We further measure the phonon number for the COM mode and the tilt mode by comparing the blue and red sideband dynamics under a narrow Raman laser on the edge ion. With a weak driving strength, the off-resonant coupling to all the other phonon modes can be neglected and effectively we get a Jaynes-Cummings model. We can thus estimate the average phonon number by comparing the blue and the red motional sidebands \cite{leibfried2003quantum} as shown in Fig.~\ref{fig:phonon}. This gives us $\bar{n}=0.9$ for the COM mode and $\bar{n}=0.5$ for the tilt mode.

We attribute the higher phonon number of the COM and the tilt modes to their higher heating rates.
For the COM mode, it is well-known that the heating rate will scale linearly with the ion number if the heating is mainly caused by a uniform electric field noise across the ion chain. In practice, the electric field noise will have slow spatial variations, so that we can expect decreasing effects on phonon modes with lower frequencies and smaller spatial wavelengths \cite{PhysRevA.100.022332}. In this experiment, due to the misalignment of the trap electrode, we expect a relatively large spatial variation of the electric field, which contributes to the heating rate of the tilt mode.

\section{Quantum state tomography}
To compare the dynamics of the spin-boson model from the initial states $|0\rangle$ and $|1\rangle$, we simply measure the final spin state in the $|0\rangle$ and $|1\rangle$ basis ($\sigma_z$). To compare the dynamics from initial states in the $|+\rangle$ and $|-\rangle$ superposition basis ($\sigma_x$), we further need to perform quantum state tomography for the density matrix of the spin and to compute the trace distance $D(\rho_+, \rho_-) = \frac{1}{2} \operatorname{Tr}|\rho_+ - \rho_-|$.

To reconstruct the spin density matrix, we measure the expectation values of $\sigma_x$, $\sigma_y$ and $\sigma_z$ to get $\rho=(I+\sum_i \langle\sigma_i\rangle \sigma_i)/2$. Among these observables, $\sigma_z$ can directly be measured in the $|0\rangle$ and $|1\rangle$ basis. To measure $\sigma_x$, we apply a $\pi/2$ pulse on the target ion using the Raman laser with the same phase as the one to prepare the initial $|+\rangle$ and $|-\rangle$ states. As for $\sigma_y$, we shift the phase of the $\pi/2$ pulse by an additional $\pi/2$.

\section{Numerical simulation}
In this section we describe how we numerically simulate the dynamics of the spin-boson model with one spin and $N$ bosonic modes. Due to the exponential increase in the Hilbert space dimension, even if we truncate each collective mode to a moderate phonon number of $n_{\mathrm{cut}}=3$, the total dimension of the system can still scale as $2(n_{\mathrm{cut}}+1)^N=2^{41}\approx 2\times 10^{12}$ for $N=20$ modes. However, for our typical phonon number of $0.3$-$0.9$ for each mode, there is still non-negligible probability for them to have a phonon number above this truncation.

Inspired by the fact that our driving laser is only able to couple to a few adjacent phonon modes satisfying $|2\eta_k b_k \Omega| \gtrsim |\Delta-\omega_k|$, we can discard the phonon modes that are not significantly excited by the spin-phonon interaction. Specifically, we give a weight $[2\eta_k b_k \Omega/(\Delta-\omega_k)]^2$ to each mode $k$ and only keep the $K$ modes with the largest weight. Here $K$ is a hyperparameter. In practice, we can gradually increase $K$ and check if the result has converged. For the numerical simulation relevant to this experiment, often we find a truncation between 5 and 10 to be sufficient.

Next, we observe that the Hamiltonian Eq.~(1) of the main text conserves the total particle number. Therefore, given an initial Fock state of one spin and $K$ phonon modes, its time evolution is restricted in a subspace of dimension
\begin{equation}
D=C(M+K-1,K-1)+C(M+K-2,K-1)
\end{equation}
where $M$ is the total excitation number of the initial Fock state, and $C(n,m)\equiv n!/[m!(n-m)!]$ is the combination number to choose $m$ items from $n$ elements. This allows us to express the Hamiltonian in a more compact and efficient way. For an initial superposition state of the spin, we can compute the dynamics in the two subspaces individually and then combine them together coherently to obtain the reduced density matrix of the spin.

Finally, we assume that initially the phonon modes are in thermal states with the average phonon number given by the measured value in the experiment. Specifically, for $N=20$ ions we set the average phonon number for the COM mode to be 0.9, that of the tilt mode to be 0.5, and that of all the remaining modes to be 0.3. For $N=10$ ions we have better sideband cooling so that we simply choose $\bar{n}=0.2$ for all the modes. We randomly generate $S=100$ trials for the initial phonon number distribution and average over them to obtain the numerical results for the spin dynamics.

%